# (p,T,H) phase diagram of Heavy Fermion systems : Some systematics and some surprises from ytterbium.

D. Braithwaite[1*], A. Fernandez-Pañella[1], E. Colombier[1], B. Salce[1], G. Knebel[1], G. Lapertot[1], V. Balédent[2], J.-P. Rueff[2,4], L. Paolasini[3], R. Verbeni[3], and J, Flouquet[1]

[1]SPSMS UMR-E CEA / UJF-Grenoble 1, INAC, 38054 Grenoble, France

[2]Synchrotron SOLEIL, L'Orme des Merisiers, Saint-Aubin, BP 48, 91192 Gif-sur-Yvette Cedex, France

[3]ESRF, 6 Rue Jules Horowitz, BP 220, 38043 Grenoble Cedex, France

[4]Laboratoire de Chimie Physique–Matière et Rayonnement, CNRS-UMR 7614, UPMC, 75005, Paris.

Abstract

Pressure is the cleanest way to tune heavy fermion systems to a quantum phase transition in order to study the rich physics and competing phases, and the comparison between ytterbium and cerium systems is particularly fruitful. We briefly review the mechanisms in play and show some examples of expected and unexpected behaviour. We emphasize the importance of the valence changes under pressure and show how modern synchrotron techniques can accurately determine this, including at low temperature.

## Introduction

In Heavy Fermion systems the 4f or 5f electrons of a rare earth (RE) or actinide element diluted in a metallic lattice, hybridize with the conduction electrons leading to a strong renormalisation of the microscopic parameters and to a large effective mass of the carriers. Although they were discovered over 30 years ago, they still play a central role in condensed matter physics today. Heavy fermion behaviour has been found in systems containing several different RE elements, particularly cerium and ytterbium, but also with samarium and praseodymium, as well as in some uranium systems. The emphasis of research has now shifted from the simple heavy fermion state to the rich phase diagrams and competing ground states, and the excitations associated with the corresponding quantum phase transitions. One of the major motivations for the study of heavy fermion systems is that a quantum phase transition between magnetic and non-magnetic ground states can be relatively easily attained by tuning the competing interactions, and high pressure is the cleanest way to do this. Many studies on cerium based systems now show that when a magnetic Quantum Critical Point (QCP) is approached in this way, several novel phenomena are found, including deviations from Fermi liquid theory in the electronic terms of resistivity and specific heat, and most spectacularly, unconventional superconductivity[1].

The comparison between Ce and Yb is a particularly fruitful path for unravelling the details of the complex behaviours found. However there are to date far fewer studies on Yb based compounds. This is partly for practical reasons. Preparing very pure single crystals of the ytterbium based systems has often proved difficult, and higher pressures are often necessary.

Both Ce and Yb are characterised by the RE ion fluctuating between a trivalent magnetic state, and a non magnetic state corresponding to an empty 4f shell for Ce, and a full (4f14) state for Yb. In both cases pressure tends to increase the delocalisation of the 4f electrons, thus driving Ce towards a non magnetic (4f$^0$) state whereas in Yb pressure will favour the magnetic Yb$^{3+}$ (4f$^{13}$) state. Thus Yb is often considered to be the "hole" equivalent of cerium. As pointed out in a review by Flouquet and Harima[2] there are however important differences between Yb and Ce, namely the deeper localisation of the 4f electrons in Yb leading to a much narrower width of the 4f level D, and stronger spin orbit coupling in Yb than in Ce. The important result is a different hierarchy of the relevant energy scales, allowing formation of the heavy fermion state for a much wider range of the valence in Yb than in Ce.

The way pressure acts to tune the system to its magnetic instability is also somewhat more more subtle than the above-mentioned picture of driving the RE ion from a magnetic to a non-magnetic valence state. The usual treatment is based on Doniach's diagram[3]. The ground state of a Kondo lattice system will depend on the relative strength of two competing phenomena, i.e. the Kondo effect which through screening of the local moments by the conduction electrons will favour a non magnetic state, and the RKKY interaction which gives the coupling between adjacent moments through the conduction electrons, and favours magnetic order. Both these effects depend on the coupling J between the local moment and the conduction electrons, but with different dependences:    $T_{RKKY} \sim J^2$ , and $T_K \sim \exp(-1/J)$. At small J, the RKKY exchange is dominant and the system orders magnetically. For intermediate J, $T_{RKKY}$ and $T_K$ are of comparable strength, magnetic order may occur with increasingly screened moments. Magnetism is suppressed with further increase of J.

The link between pressure and the coupling parameter J, as well as with the valence change between magnetic and non-magnetic states described above, is non trivial. The Doniach picture does not explicitly refer to the valence change mechanism, in fact on the contrary the original Doniach paper states precisely that the effects of non-integer valence states have not been taken into account. In heavy fermion systems the remarkable properties stem precisely from this non integer occupation, nevertheless the Doniach picture can still give a good picture of the main physics. The coupling J is proportional to $D/|e_f - E_F|$, where D is the width of the 4f level, $e_f$ and $E_F$ the 4f and the Fermi energies respectively. For cerium systems the translation of the effect of pressure is clear. Under pressure D will increase, the energy of the $4f^1$ level will increase moving towards the Fermi energy, so $|e_f - E_F|$ will decrease. Both these effects combine to drive the system towards increasing J in the Doniach diagram, leading to the suppression of magnetic order, even though the Ce valence may still be close to the magnetic $Ce^{3+}$. For ytterbium the situation is less straightforward. The dominant effect will be the change of $|e_f - E_F|$. Pressure will still increase the energy of the 4f level but $e_f$ is now the energy of the 4f hole lying above $E_F$. $|e_f - E_F|$ will therefore increase with pressure, driving the system towards decreasing J in the Doniach diagram, and so inducing magnetic order. $e_f$ is obviously strongly dependent on the Yb valence, so establishing the link between the Doniach picture and the valence change. However for ytterbium D will also increase with pressure. Thus the terms in J may be in competition, leading to a less effective pressure effect in ytterbium systems compared to cerium. Finally if the valence change of Yb with pressure becomes zero or very small, for example when the $Yb^{3+}$ state is reached, although $e_f$ may continue to increase with pressure we expect this variation to be much weaker. In principle it could even be possible to find an ytterbium system where J actually increases with increasing pressure. To illustrate these points we present measurements on 3 different Yb systems: $YbCu_2Si_2$, $YbAl_3$, and $YbNi_3Al_9$.

## Experimental

Flux growth is often suited to intermetallic Yb systems. Single crystals of $YbCu_2Si_2$ were grown by an indium flux method. $YbAl_3$ [4] and $YbNi_3Al_9$ were grown from Al flux. Measurements under pressure on $YbCu_2Si_2$ and $YbNi_3Al_9$ were performed in a diamond anvil cell, with argon pressure transmitting medium for good hydrostaticity. Pressure was determined from the ruby fluorescence by placing some ruby chips in the pressure chamber. In the case of $YbAl_3$ measurements were also performed in a Bridgman cell with steatite transmitting medium for very high pressures (24 GPa) and a liquid (Fluorinert) medium up to 6 GPa. In the diamond cell we performed resistivity measurements with 4 gold wires microwelded to the sample[5]. We also performed microcalorimetry measurements with a Au/AuFe thermocouple attached to the sample[6], and a.c. susceptibility measurements by inserting a micro-coil into the sample chamber[7]. In the Bridgman set-ups only resistivity was measured. For the diamond anvil cell pressure could be changed in-situ at low temperature using a helium bellows system and a force amplification mechanism[8], in a standard 4He cryostat, and very recently in a dilution fridge where measurements down to 50 mK have been performed.

## Resuts and discussion

$YbCu_2Si_2$ is the ideal well-behaved prototype Yb system. It has the tetragonal $ThCr_2Si_2$ structure At ambient pressure it is a moderate Heavy Fermion system ($g=135$ mJ/mol.$K^2$) and an intermediate valence of about 2.85. Resistivity measurements under high pressure first showed that the Kondo temperature, which can be roughly extracted from the broad maximum in the resistivity, decreases with pressure as expected. At pressures above 8 GPa a distinctive anomaly appears which was attributed to magnetic order. Further studies using Mössbauer spectroscopy, micro-calorimetry[9], ac-susceptibility have all confirmed the presence of long range magnetic order at high pressure (see figure 1).

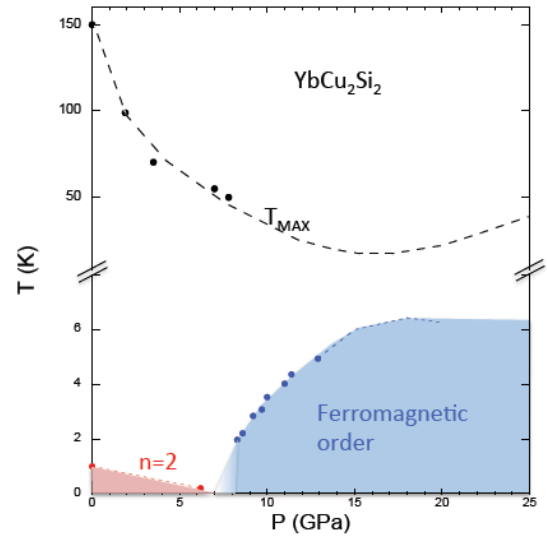

Figure 1. Phase diagram of $YbCu_2Si_2$, adapted from references[10-13]. The red shaded area shows the limit of the $T^2$ Fermi liquid behavior of the resistivity.

Subsequently we have performed resistivity experiments with a new generation of higher quality single crystals, with greatly reduced residual resistivity (RRR up to 200) and improved hydrostatic conditions. These confirm the findings of the initial study, that the phase diagram is at least qualitatively in agreement with spin fluctuation models: The temperature below which a quadratic temperature dependence of the resistivity is found ($r = r_0 + AT^2$), decreases with a concomitant large increase (2 orders of magnitude) of A[9]. So far this conforts the picture of the mirror image of a cerium system phase diagram. In fact $YbCu_2Si_2$ could be compared in many aspects to its cerium counterpart $CeCu_2Si_2$. However there are several significant differences. The first surprise was that despite a considerable experimental effort to measure very close to the critical pressure and at low temperature it has so far been impossible to detect a phase transition below 1K. Magnetism seems to occur suddenly, at a pressure of 7.5-8GPa, and a temperature of 1.2-1.3 K. The possibility of a first order transition was first proposed from the Mössbauer results where the coexistence of a magnetic and nonmagnetic component was found over a pressure range of several GPa[13], and this seems to be confirmed from the phase diagram. The second surprise comes from the a.c. susceptibilty measurement which show that the magnetic order which occurs is ferromagnetic[12]. Actually this is not completely surprising as several other Yb systems order ferromagnetically at ambient pressure[14], and there is indirect evidence for ferromagnetism under pressure in at least 2 other systems[15]. It seems that the presence of strong ferromagnetic correlations may be a general trend in Yb systems, on the contrary to cerium. A possible explanation was suggested based on the larger valence changes expected, and the strong effect of magnetic field on a possible valence transition[12,16].

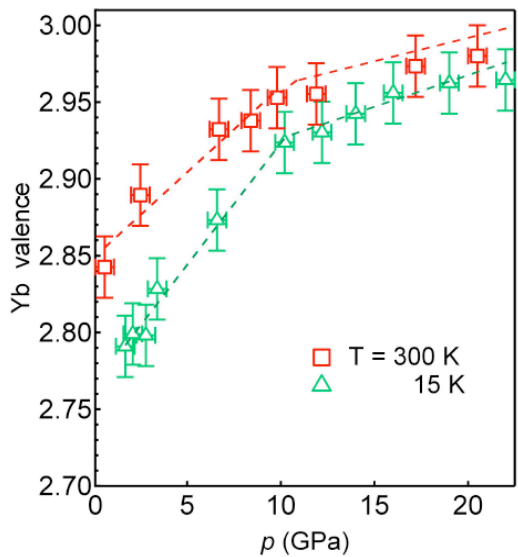

Figure 2. Variation of Yb valence in $YbCu_2Si_2$ under pressure at room temperature and at low temperature measured by RIXS. Adapted from[17]

From all these results it is clear that the next step to understand the phase diagram of $YbCu_2Si_2$ is to clarify how the Yb valence changes with pressure. For this resonant inelastic xray spectroscopy is now a powerful and accurate tool. However in order to get precise information about valence changes it is necessary to combine extreme conditions of high pressure and low temperature. Such an experiment has recently been performed at the ESRF[17]. The resulting valence change, measured at low temperature close to the magnetic ordering temperature is shown in figure 2. The remarkable feature is the clear change of slope at a pressure close to the critical pressure where magnetism occurs. However magnetic order clearly occurs before the $Yb^{3+}$ state is realized. The Yb valence continues to increase slightly over the whole pressure range explored, thus conforting the Doniach description. The interesting question is what will happen once the trivalent state is reached. Indeed the resistivity measurements show almost no increase of the ordering temperature at very high pressure. In fact Winkelman et al.[13] even found a slight decrease above 20GPa, though this was within the error bars. Interestingly the broad maximum, taken as an indication of the Kondo temperature, was also found to move towards higher temperatures pressures above about 18GPa. Although as pointed out by the authors this is not a completely reliable indication of the behaviour of the Kondo temperature as it also depends on other effects, especially the crystal field. Nevertheless the possibility exists that the pressure effect on J is reversed at high pressure. Unfortunately the very high pressures necessary to study this effect in detail make this rather difficult. It is probably more productive to search for similar effects in other systems, but at lower pressure.

$YbAl_3$ should offer a nice possibility to study a system situated slightly further from the critical pressure than $YbCu_2Si_2$. It has a high symmetry cubic structure. At ambient pressure the Yb valence is about 2.75, the effective mass is moderately enhanced (g=50 mJ/mol.K$^2$), and estimations of the Kondo temperature give a rather high value (400-600K)[4,18]. On the other hand it is a softer material than $YbCu_2Si_2$, almost 3 times more compressible[19,20]. We naturally expected that with the application of pressure we would be able to attain the critical point and induce magnetic order.

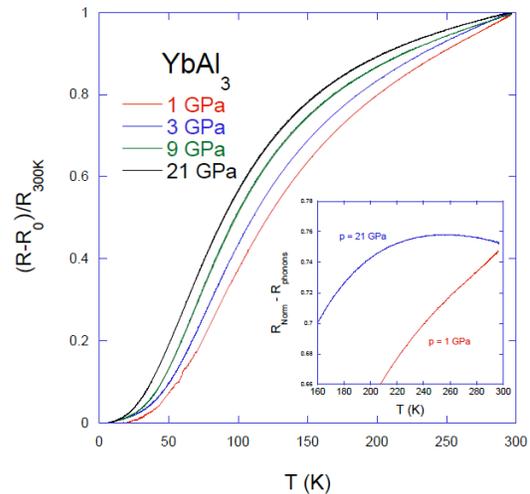

Figure 3. Normalized resistivity versus temperature for $YbAl_3$ under pressure. The inset shows the result after subtraction of the approximate phonon contribution showing the appearance of a broad maximum significant of the decrease of the Kondo temperature.

In fig 3 we show resistivity curves measured in the Bridgman cell up to 21 GPa. The low pressure curve shows no maximum in the resistivity, in agreement with the high Kondo temperature. Under pressure the magnetic contribution clearly increases and shifts to lower temperatures. An analysis by subtracting the estimated phonon contribution showed that a maximum appears for pressures above 6 GPa, and shifts to about 220K at the maximum pressure measured. A more accurate measurement performed with liquid medium found an increase in the A coefficient by a factor 2 up to 6GPa. These effects go in the expected direction, but are surprisingly weak. Again we turn to the pressure effect of the Yb valence to understand.

Unfortunately no studies have been performed at low temperature, however the pressure dependence has been measured at 300K[19]. They find that the valence indeed increases, but at a much slower rate than in $YbCu_2Si_2$. At the highest pressure measured (38 GPa) the valence reaches 2.93, even though this pressure corresponds to an almost 25% volume decrease. This result confirms the importance of the valence state in order to describe the phase diagram. For the anecdote our resistivity results were not published at the time as the result seemed rather disappointing, but in fact this absence of effect could be an important element in understanding the global picture for ytterbium systems.

Finally we search for an example of an Yb system where pressure has the reverse effect. A very recent study on $YbNi_3Al_9$[21] found that magnetic order occurs at ambient pressure, and that the ordering temperature decreases with increasing pressure. This compound has lower symmetry than the previous examples, having the trigonal $ErNi_3Al_9$ - type structureWe decided to check this surprising effect. New single crystals were synthesized from Al flux. Heat capacity measurements at ambient pressure showed a sharp feature at 3.2 K confirming the presence of magnetic order. To test the bulk nature of the observed pressure dependence we performed microcalorimetry measurements in the diamond anvil cell up to 10 GPa.

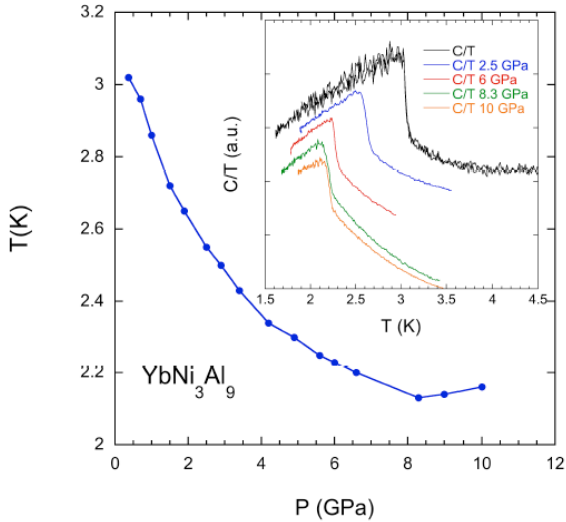

Figure 4. Phase diagram of single crystal $YbNi_3Al_9$ determined from calorimetry measurements. The inset shows the C/T curves for selected pressures.

The resulting phase diagram confirms the decrease of the ordering temperature with quite a strong effect (-0.2K/GPa). However the expected critical point was not reached as above 4 GPa the pressure effect flattens out. In fact the highest pressure points the ordering temperature seems to increase slightly, although this is within the experimental uncertainty. Apart from the disappointment of not reaching a QCP, this result is quite surprising. Xray absorption and magnetic measurements show a valence state close to $Yb^{3+}$ in this system[22]. It could therefore be a good candidate for a system where the valence changes very little with pressure, and J is dominated by the competing effects leading to the suppression of magnetic order with pressure. However in this case there is no reason for the pressure effect to weaken at higher pressures, except that as usual if the total exchange is the result of several competing phenomena the result can be rather unpredictable. It was also suggested that the order could change from antiferromagnetic to ferromagnetic at about 4 GPa[21]. Further experiments under pressure to determine how the Kondo temperature and the valence evolve with pressure are definitely desirable in this system.

|  | $T_K$ | $T_{CEF}$ | Yb valence |
|---|---|---|---|
| $YbCu_2Si_2$[23] | 40K-60K | 100K-150K | 2.85 |
| $YbAl_3$[18] | 600K | 50K ? | 2.75 |
| $YbNi_3Al_9$[24] | 5K | 18K | 3 |

Table 1. Summary of relative energy scales and Yb valence for the 3 examples. For $YbAl_3$ a direct determination of the CEF is not available, but it is likely that the low energy scale found for the onset of coherence[18] corresponds to the lifting of the degeneracy by the CEF.

From these examples we see that the simple Doniach picture presented is probably not sufficient to describe the variety of behaviours found. This is not really surprising as with increasingly high pressures other factors will come into play. At some point the direct coupling will even become important. But perhaps the most important initial factor to take into account is the competition between Kondo and crystal electric field (CEF) effects. In table 1 we summarize the values at ambient pressure. The particularity of $YbCu_2Si_2$ is that the CEF energy is rather high leading to a clean situation under pressure where the CEF can lift the degeneracy well above the Kondo temperature. The case of $YbNi_3Al_9$ is quite opposite. The $Yb^{3+}$ ion is surrounded by a lot of atoms, and the CEF is rather low, comparable to $T_K$. This situation can give rise to more complex behaviour.

## Conclusions

We show that ytterbium heavy fermion systems under pressure while generally showing the expected "4f hole analogue" behaviour as compared to cerium, also display specificities due to the more localized nature of the 4f electrons, and the competition of different effects on applying pressure. Whereas an understanding of the cerium heavy fermion phase diagrams more or less exists, ytterbium undoubtedly deserves more attention and probably reserves more surprises. The knowledge of the valence variation is an important parameter to take into account.

## Acknowledgements

We thank T. Ebihara for providing the $YbAl_3$ crystals, H. Harima and D. Jaccard for useful discussions. This work was partly supported by the french agency ANR through the project PRINCESS.